\documentstyle[preprint,epsfig,aps]{revtex}

\begin{document}

\centerline{\bf Consequences of parton's saturation and string's
percolation} 
\centerline{\bf on the developments of cosmic ray showers}

\centerline{ C. Pajares, D. Sousa, and R.A.~V\'azquez}
 
\centerline{\it Departamento de F\'\i sica de Part\'\i culas, Universidade de
Santiago}
\centerline{\it E-15706 Santiago de Compostela, Spain}

Pacs Numbers 13.85.Tp, 13.85.-t, 96.40.De

\begin{abstract}
At high gluon or string densities, gluons' saturation or the
strong interaction among strings, either forming colour ropes or 
giving rise to string's percolation, induces a strong suppression in the
particle multiplicities produced at high energy. This suppression
implies important modifications on cosmic ray shower
development. In particular, it is shown that it affects the
depth of maximum, the elongation rate, and the behaviour of
the number of muons at energies around $10^{17}$--$10^{18}$
eV. The existing cosmic ray data point out in the same direction.
\end{abstract}

One of the most crucial astrophysical issues of the highest energy
cosmic rays (above $\sim 10^{17}$ eV) is that of their composition. 
This problem is linked to the identification of the origin and 
possible sources of these cosmic rays. Current theoretical models 
expect a transition from galactic to extragalactic or galactic halo 
sources near the region of the ankle which leads to the usual 
expectation of the changing of composition from heavy to light 
elements.

Experimentally, measuring the composition at these energies is a
challenging task. The very low fluxes involved imply that one has to 
rely on indirect measurements which depend on simulations of the 
development of cosmic ray cascades in the atmosphere. These, in 
turn, 
are model dependent and, specifically, depend on extrapolations of 
hadronic models to energies and kinematical regions never measured
in the laboratory. There is, therefore, some degree of 
uncertainty in the shower development and one may ask what is the
effect of this uncertainty in the reconstruction of shower 
parameters, mainly total energy and mass composition.

To avoid this problem, experimental groups have concentrated on 
observables which are expected to be more or less independent of the 
hadronic model used, or which have its dependence under theoretical 
control. These parameters include the maximum of the cascade, 
$X_{\rm{max}}$, the slope parameter, 
$\beta = d \log(\rho_\mu(600))/d \log(E)$, where $\rho_\mu(600)$ is 
the muon density at 600 m from the core, and the elongation rate,
$D_{10}= d X_{\rm{max}}/d \log_{10}(E)$. Other parameters 
have been less frequently used, see Ref.\cite{Watson} for a general
review.

Several experimental groups have measured the cosmic ray spectrum
and mass composition in the ankle region and beyond using the
above mentioned parameters 
\cite{FlysEye,Hayashida_1,Nagano,hiresmia,Yakutsk} see 
also\cite{Watson}. The results on 
mass composition are inconclusive. Fly's Eye \cite{FlysEye} observe 
a change on the
slope of $X_{\rm{max}}$ in the region around $3 \times 10^{17}$
eV, which is interpreted as a change on the composition from heavy
(iron) dominated to light (proton) dominated. However, AGASA
\cite{Hayashida_1} measures a muon component and $\beta$ parameter
consistent with iron on that region. Although some part of the
discrepancy between AGASA and Fly's Eye may be due to the use of
different hadronic models \cite{Dawson}, as pointed out by Nagano
{\it et al.}  \cite{Nagano} the important issue is that AGASA sees
no significant change on the muon component of showers all along
the ankle region and beyond, from $10^{16.5}$ eV to $10^{19.5}$
eV, and thus no strong change on composition is inferred.
HIRES and MIA collaborations \cite{hiresmia} have jointly
measured both the $X_{\rm{max}}$ and the $\beta$ parameter. They
observe a strong change of $X_{\rm{max}}$ with energy, which
implies a large elongation rate, $D_{10}= 95$ gr/cm$^2$ and on the
other hand they see no change on the slope of the muonic component, 
$\beta = 0.73$, measured which is broadly compatible with the AGASA
observations. HIRES and MIA however have measured these parameters
in a narrow range of energy, from $10^{17}$ to $10^{18}$ eV and 
with low statistics, only during a limited exposure.

In this paper we will show that under rather general conditions
a change on the hadronic interactions at the energies of 
interest is expected, which may have important consequences for the 
interpretation of cosmic ray data.
Whether this change is enough or not to produce the
observed changes on the cosmic ray data we can not tell at present. 
On the other hand we can state the necessary conditions for this 
change to explain the observed data: i) There should be an abrupt
change on the hadronic interactions at the observed energy 
$E_{\rm{{\small lab}}} \sim 5 \times 10^{17}$ eV for Fe--Air 
collisions.
This corresponds to a CM energy of $\sim 4200$~AGeV 
and a density of gluons of $\sim 9$~fm$^{-3}$ ii) At this energy 
the slope of the growth of the multiplicity with energy should vary
from $\sim 0.24$ to a essentially flat $\lesssim 0.09$. 
If i) and ii) are verified then there is no additional need for a 
change on composition to explain the data. 

It is important to point out that, although the change
on the multiplicity may or may not be enough to produce the observed
results, some effect should always be present and should be taken into
account in any realistic simulation of cosmic ray showers. Currently
no Monte Carlo code for cosmic ray showers has yet been 
implemented with these effects\footnote{Sibyll version 2.0 
\cite{Sibyll} incorporates some 
shadowing effect. However this was done for $pp$ collisions only 
and does not affect to our reasoning below.}

In the last years, a wealth of data coming from HERA
and the heavy ion SPS experiments have risen questions about the
behaviour of the hadronic interactions at very high energy.
We may consider perturbative, gluons, or non perturbative, strings,
as the fundamental variables of our description, 
At high gluon density, the saturation of gluons\cite{QCD} and/or a
strong jet shadowing\cite{Eskola} are expected. In the case of high
string density we expect the fusion of strings\cite{fusion} or
colour rope\cite{sorge}, and probably, above a critical string
density the percolation of strings\cite{percolation} and
the formation of quark gluon plasma at the nuclear scale are
expected.

One general feature of all these hadronic phenomena is the strong
suppression of particle multiplicity compared to the multiplicity
expected in their absence. Namely, for central Pb--Pb collisions
the charged particle multiplicity expectations in the central 
rapidity region changes between 1500 (7500) for the 
relativistic heavy ion collider, RHIC, (the large hadron collider, 
LHC) for models that do not include these effects to 900 (3000)
when they are included\cite{Armesto}.
As a framework we will use the quark gluon 
string model (QGS)\cite{QGSM}, a modified version of the Dual Parton
Model\cite{DPM}. The model is based on the large $N$ expansion of
QCD but it is largely phenomenological and describes most of the
soft hadronic physics rahter well. Inclusion of hard, perturbative, 
physics has been done in various ways. In the quark--gluon string 
model, multiparticle production is related to the interchange of 
multiple strings which break and subsequently hadronize.

In this model one can most easily understand the expected changes
on the behavior of hadronic collisions at high energies.  It is
more convenient to work in the plane transverse to the
collision. In this plane, strings are seen as small circles of
fixed radius, $r$. As the energy increases, the number of strings
interchanged increases and the total area occupied by strings
increase. At high energy, strings start to overlap and fuse
together. For high enough string density, $n_c$, strings may
percolate in a second order phase transition, {\it i.e.}
continuous paths of strings are formed in the collision
area. Since the number of independent strings is reduced after the
fusion one expects a depletion on the number of particles
produced, i.e. a reduction on the multiplicity.  

In the QGS the multiplicity grows with energy
as $n(s)\sim s^\Delta$, where $\Delta$ is related to the intercept of 
the soft pomeron\cite{QGSM}\footnote{Here we consider minimum bias 
events, which are relevant for cosmic ray experiments. The parameter 
$\Delta$ may depend on the centrality of the collision}. 
In the case of percolation,
the reduction of multiplicity is given by \cite{Braun,Armesto} 
\begin{equation}
n'(s) = n(s) \sqrt( F(\eta)),
\end{equation}
where 
\begin{equation}
F(\eta) = \frac{1- e^{-\eta}}{\eta},
\end{equation}
and the parameter $\eta$ is the fraction of the total area occupied
by strings
\begin{equation}
\eta= \frac{\pi r^2 N_s}{\pi R^2}.
\label{eq:eta}
\end{equation}
Here $N_s$ is the number of strings produced in the collision, $r$
is the string's transverse size, and $R$ is the total collision area.
$N_s$ grows with energy as $N_s \sim s^{\Delta'}$, where 
$\Delta'$ is the intercept of the soft pomeron\cite{QGSM}. Therefore 
at large $\eta$ the total multiplicity grows with energy as
\begin{equation}
n'(s) \sim s^{\Delta- \Delta'/2}.
\label{eq:fusion}
\end{equation}

This reduction of multiplicity is not
exclusive of the percolation of strings. For instance, in 
perturbative QCD a reduction in the number of 
jets produced as the energy increases is also expected\cite{Eskola}.
At high energy the number of jets produced grows with energy as 
$n(s,p_t^2)\sim (s/(4p_t^2))^{\Delta_H}$, where $\Delta_H$ is the 
intercept of the hard pomeron and $p_t$ is the transverse momentum 
of the jet. A (mini)jet occupies a transverse area of order 
$\pi/p_t^2$, since the number of jets increases rapidly with energy, 
saturation occurs when the area occupied by the jets equals the 
total transverse area \cite{Eskola,Armesto}:
\begin{equation}
\frac{n(s,p_t^2) \pi/p_t^2}{\pi R^2} =  1,  
\end{equation}
which implies
\begin{equation}
n(s,p_t^2) \sim s^{\frac{\Delta_H}{1+\Delta_H}}.
\label{eq:shadowing}
\end{equation}

Both Eqs.(\ref{eq:fusion},\ref{eq:shadowing}) have been checked
directly in Monte Carlo simulation \cite{fusion,Eskola}.
Surprisingly, for nucleus--nucleus collisions the reduction in the 
power of multiplicity growth with energy is of the same order 
both for the case of string fusion and of shadowing. The power 
changes from $\sim 0.24$ to $\sim 0.19$. In general parton 
saturation, shadowing, string fusion, or percolation  will 
produce the effect of reduction of the multiplicity although we 
expect the degree of this reduction to be model dependent. 

For cosmic ray showers, the rate of change of the multiplicity
with energy is directly related to the change of the elongation rate. 
This has been known for a long time as the elongation rate 
theorem \cite{Linsley}. The elongation rate theorem can be deduced 
easily, as it follows from a scaling argument. Let 
$X_{\rm{max}}(E)$ be the maximum depth of the shower 
produced by a primary of energy $E$. On average, the first 
interaction occurs at depth $\lambda$, the mean
free path of the initial particle. In this first interaction the initial 
particle splits into $n(E)$ particles each carrying an average energy
$E/n(E)$. Therefore, we have
\begin{equation}
X_{\rm{max}}(E)= \lambda + X_{\rm{max}}(E/n(E)).
\end{equation}
Assuming that $X_{\rm{max}}(E)$ depends logarithmically on 
energy we get
\begin{equation}
X_{\rm{max}}(E)= A \log_{10}(E/n(E)) + B,
\end{equation}
where $A=X_0 \ln 10$ and $B$ are constants. $X_0=37 $ gr/cm$^2$ is 
the electromagnetic radiation length. If we now assume that $n(E) 
\propto E^\Delta$, we get
\begin{equation}
X_{\rm{max}}(E)= A (1-\Delta) \log_{10}(E) + B'.
\end{equation}
This is the elongation rate theorem. We can now directly read the 
elongation rate from the above equation $D_{10}= A (1- \Delta)$. As
stated previously, a change in the multiplicity growth with energy
implies a change in the elongation rate. 

\begin{figure}
\epsfxsize=10cm
\begin{center}
\mbox{\epsfig{file=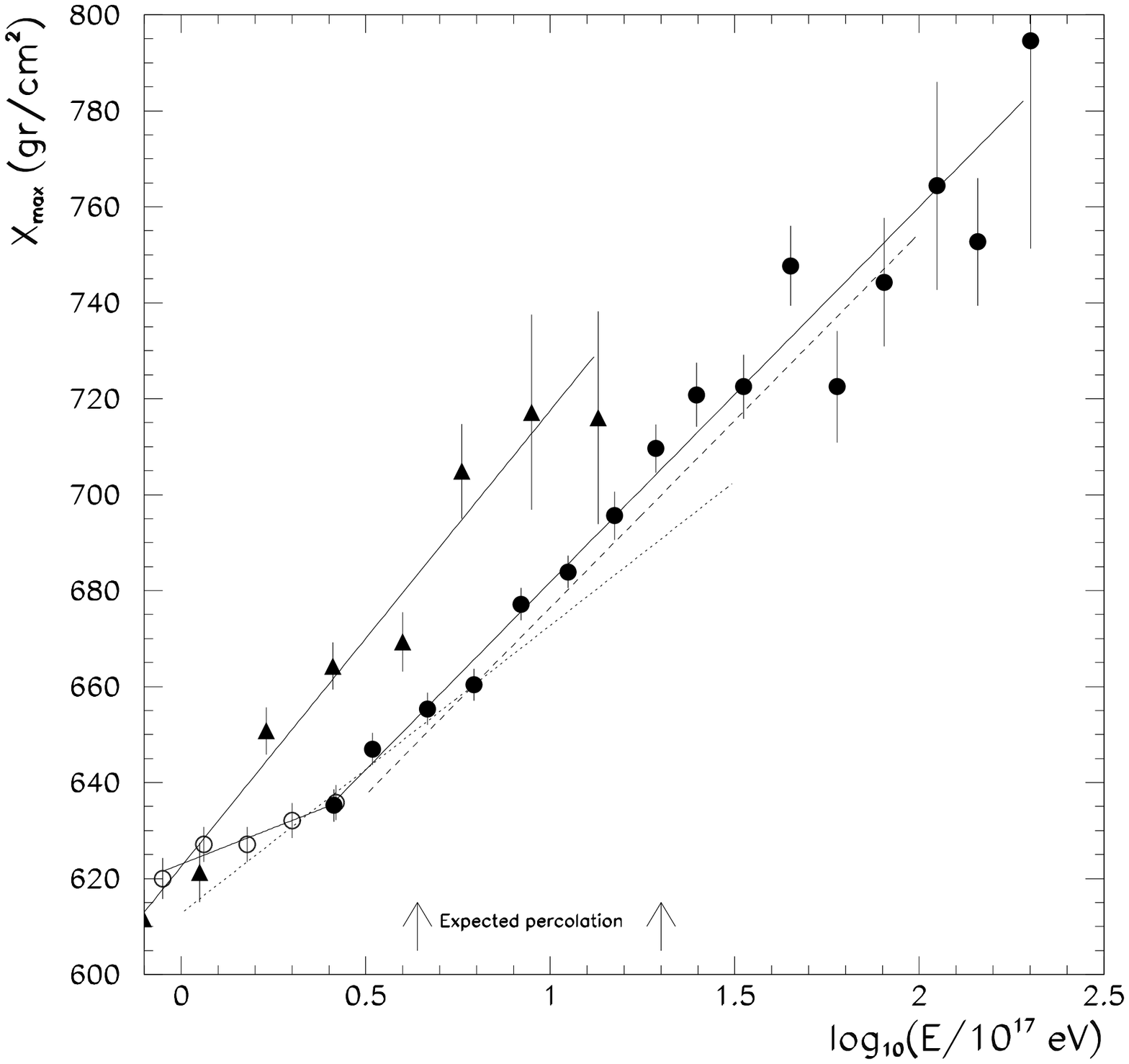}}
\end{center}
\caption{Depth of shower maximum as a function of the logarithm of 
the primary energy as measured by Fly's Eye (circles) and HIRES--MIA
(triangles). Full lines are fits to the data. Dashed and dotted
lines are our prediction for a strong reduction on the
multiplicity at an energy of $\sim 6 \times 10^{17}$ eV. Arrows
mark the expected region where percolation occurs for Fe--Air
collisions in the string fusion model.}
\label{flyseye:fig}
\end{figure}

In Fig.(\ref{flyseye:fig}) we show $X_{\rm{max}}$ as a function 
of energy for the Fly's Eye and HIRES-MIA experiments. Data have been 
taken from references \cite{FlysEye,hiresmia}. The errors shown are 
only statistical. An additional systematic error of 
$\sim 20$ gr/cm$^2$ must be included in the data. The dash line 
represents our calculation for the slope parameter, 
$D_{10} = 65 $ gr/cm$^2$,($\Delta=0.24$ from our simulations) for 
Fe--Air collision without fusion, normalized with the data at 
$6 \times 10^{17}$ eV. The dotted curve 
has a slope parameter $D_{10} = 78 $ gr/cm$^2$, 
which would imply a maximum reduction in the slope of growing of 
multiplicities: from $\Delta \sim 0.24 $ to $\Delta \sim 0.09$. 
The data from HIRES--MIA
is not completely consistent with the Fly's Eye data. Statistical 
uncertainties are larger. The elongation rate obtained by the 
HIRES-MIA collaboration is very large, $95 $ gr/cm$^2$. Notice that 
the elongation rate theorem predicts an elongation rate always less 
than 85 gr/cm$^2$, an elongation rate larger would imply 
multiplicities {\it decreasing} with energy. Therefore, if 
the HIRES-MIA result is confirmed, a change of the composition is
necessary to explain the data. In the figure we also show the energy 
region in which a phase transition is expected in Fe--Air collision
using the string fusion model\cite{fusion2}. This region corresponds 
to the range obtained from percolation theory, 
$1.1 \leq \eta \leq 1.2$, where $\eta$ is given by 
Eq.(\ref{eq:eta}). 

A word of caution is necessary in the use of the elongation rate 
theorem. The elongation rate theorem is based on the
assumption that the energy is equally shared between the secondaries
in the hadronic interaction. From this assumption immediately follows
the logarithmic dependence of $X_{\rm{max}}$ on the 
multiplicity. In realistics cases this assumption does not hold and 
one has to resort to simulations since no analytical formula is
known for the $X_{\rm{max}}$. We have parameterized for a number 
of models the dependence of $X_{\rm{max}}$ on the change of 
multiplicity, see Ref.\cite{fusion2} for details. The results
of a full Monte Carlo agree with our qualitative discussion.
With this in mind we can conclude that to be able to explain the
change on the elongation rate we need a change on the slope of
the multiplicity from $\Delta \sim 0.24 $ to an essentially flat 
$\Delta \sim 0.09$. The energy region for such change must be 
around $5 \times 10^{17}$ eV. 

The experimental situation with the lateral distribution of muons 
$\rho_\mu(r)$ is clearer. Both in simulations and experiments 
it is found that the shape of the lateral distribution function for 
muons is rather independent of the primary's energy and 
composition. Therefore, at fixed distance
to the core, $r_0$, $\rho_\mu(r_0)$ is proportional to the total number of 
produced muons in the shower. Under rather general arguments this 
number scales with energy\cite{Gaisser}
\begin{equation}
\rho_\mu(r_0) \propto N_\mu = A E^\beta,
\end{equation}
where $A$ is a normalization constant and $\beta$ is the slope 
parameter. As 
mentioned previously, the slope parameter is found to be constant 
over a wide range of energies. This result is consistent with 
Yakutsk, Haverah Park \cite{Nash}, and with all the lower energy
experiments. 

It is rather simple to calculate the slope parameter, $\beta$, 
for a pionic
cascade from a scaling argument similar to the elongation rate
theorem. The number of muons is proportional to the number of 
charged pions at the maximum. The number of pions, at maximum, 
produced by a primary of energy $E_0$ is given by
\begin{equation}
N_\pi(E_0) = f_\pi \, n \; \int_0^1 dx P(x) N_\pi(x E_0),
\end{equation}
where $f_\pi=2/3$ is the charged pion fraction, $n$ the total 
pion multiplicity, and $P(x)$ is the 
probability of producing a pion with a fraction of energy $x$ 
of the primary energy. Assuming a scaling form, $N_\pi = A E^\beta$,
we get 
\begin{equation}
1 = f_\pi \, n \; \int_0^1 dx P(x) x^\beta = f_\pi \, Z(\beta),
\label{scaling}
\end{equation}
where $Z(\beta)$ is the spectrum--weighted momentum\cite{Gaisser}. 
For a given
$P(x)=1/n \, dn/dx$, the above equation gives an implicit equation
for $\beta$. It reduces to the textbook's expression if we assume
energy equipartition, {\it i.e.} $P(x)= \delta(x-1/n)$, which
gives $\beta= (1+\log(f_\pi))/\log(n) \sim 0.82$, for $n \sim 10$.
For realistic models the slope parameter $\beta$ ranges between
0.7--0.9. In Eq.(\ref{scaling}) the multiplicity enters explicitly
in the left hand of the expression but also enters implicitly since
the probability $P(x)$ must verify total probability and energy
conservation. Since $Z(\beta)$ is a monotonically decreasing
function of $\beta$ for reasonable choices of $dn/dx$, a reduction 
of multiplicity induces a reduction on $\beta$. Indeed this is what 
is observed for the $dn/dx$ calculated
for the model with and without fusion. The DPM model gives a value
of $\beta \sim 0.89$ which agrees with that of the QGSJET model
\cite{Nagano}. In the presence of fusion the slope parameter is
reduced and we get $\beta \sim $ 0.72--0.77 depending on the
specific implementation of the fusion model. This number was 
calculated both using Eq.(\ref{scaling}) and by direct calculation 
with a Monte Carlo code.

\begin{figure}
\epsfxsize=10cm
\begin{center}
\mbox{\epsfig{file=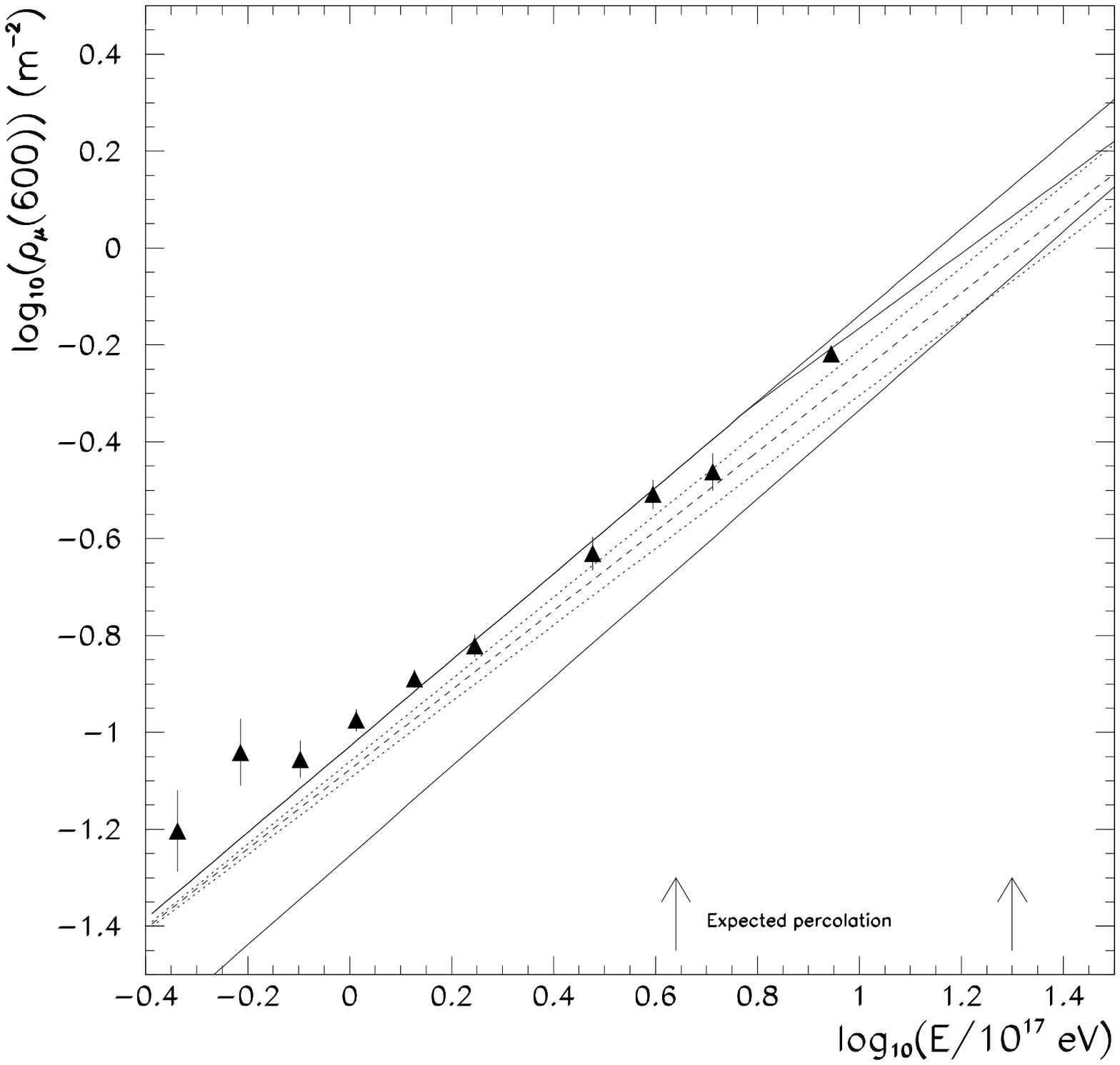}}
\end{center}
\caption{Logarithm of the muon density at 600 m as a 
function of the logarithm of the energy as measured by HIRES--MIA
(triangles) and AGASA (dashed line). Dotted lines are the errors
of the AGASA parameterization. Also shown the prediction for
the QGSJET model for pure iron (upper full line) and proton (lower
full line) and our prediction for a change of slope as given in 
the text (marked). The arrows mark the position where percolation
occurs for Fe--Air collisions in the string fusion model.}
\label{muon:fig}
\end{figure}

In Fig.(\ref{muon:fig}), we show de density of muons at 600 m,
$\rho_\mu(600)$, as a function of the energy for the AGASA
measurements given as a parameterization and the HIRES-MIA
measurements, shown as triangles. Also shown are the QGSJET results
for pure iron and proton. Notice that the slope
parameter measured by HIRES-MIA agrees with our slope parameter for
the case of fusion. Our result, a change of slope parameter from 
that of pure iron to 0.77 at the energy where percolation is 
expected, is shown for comparison only. 
We can see that again it is consistent with the data. The slope 
measured by AGASA is different from the one 
calculated for either proton or iron for the QGSJET model. 
A rapid change on composition, as suggested by the HIRES-MIA data
on $X_{\rm{max}}$, would imply a kink in the 
data for $\rho_\mu(600)$ at the same energy which is not seen.
Instead, our result points towards a mild change on the slope 
parameter, from 0.9--0.8 to 0.77 which would be hardly seen 
given the error bars in the data. 

There are a number of additional predictions in our scenario.  The
average $p_t$ in hadronic collisions should increase in the case
of string fusion by about 10 -- 20 \%. This would produce flatter
lateral distribution for the muon densities which could be
observed. A particularly well--suited place to look for this
effect would be in inclined showers. Inclined showers are composed
esentially of high energy muons, and therefore are more sensitive
to changes on the first hadronic interactions\cite{Ave}.  
Given the current
systematical and statistical errors we can not conclude that
indeed cosmic ray experiments are observing a saturation of gluons
or percolation of strings in hadronic interactions. High quality
data with large statistics, as the expected from HIRES and the
Pierre Auger observatories, are needed.  However it is suggestive
that all cosmic ray existing data are consistent with such
interpretation. RHIC and LHC will measure the total multiplicity
in the relevant energy region and ascertain whether a strong
reduction of multiplicity takes place or not.  But in any case,
even if the change of composition is real, these effects must be
taken into account in a complete simulation of cosmic ray showers.

\acknowledgements

We thank N. Armesto, E.G. Ferreiro, C. Merino, and E. Zas for 
useful discussions. This work has been partially supported by
CICYT (Spain), AEN99-0589-C02-02.

\newpage

\end{document}